# Relay Strategies Based on Cross-Determinism for the Broadcast Relay Channel

Peyman Razaghi and Giuseppe Caire
University of Southern California, Los Angeles, CA
{razaghi,caire}@usc.edu

*Abstract*—We consider a two-user Gaussian multiple-input multiple-output (MIMO) broadcast channel with a common multiple-antenna relay, and a shared digital (noiseless) link between the relay and the two destinations. For this channel, this paper introduces an asymptotically sum-capacity-achieving quantize-and-forward (QF) relay strategy. Our technique to design an asymptotically optimal relay quantizer is based on identifying a cross deterministic relation between the relay observation, the source signal, and the destination observation. In a relay channel, an approximate cross deterministic relation corresponds to an approximately deterministic relation, where the relay observation is to some extent a deterministic function of the source and destination signals. We show that cross determinism can serve as a measure for quantization penalty. By identifying an analogy between a deterministic broadcast relay channel and a Gaussian MIMO relay channel, we propose a three-stage dirty paper coding strategy, along with receiver beamforming and quantization at the relay, to asymptotically achieve an extended achievable rate region for the MIMO broadcast channel with a common multiple-antenna relay.

## I. INTRODUCTION

This paper considers a broadcast-relay channel where a common relay assists two users simultaneously, as shown in Fig. 1. We focus on simple quantize-and-forward (QF) strategies that avoid the complexity of decoding at the relay, and are able to asymptotically achieve the cut-set bound for individual rates, and for the sum capacity. The main tool advocated here to design such a QF scheme is *cross determinism*, as introduced in the following section.

### A. Cross Determinism

Consider a semi-deterministic memoryless single relay channel defined by $p(y, y_r|x)$, with a noiseless (digital) relay link of rate $R_0$ to the destination,[1] and assume that the relay observation $Y_r$ is a deterministic function of the source signal $X$ and the destination observation $Y$, i.e., $Y_r = f(X, Y)$ for some deterministic function $f(\cdot, \cdot)$. We call the relationship $Y_r = f(X, Y)$ an instance of *cross determinism*, since $X$ and $Y$ deterministically give $Y_r$. An important example of such a relay channel, is a *noise observing* relay channel where the

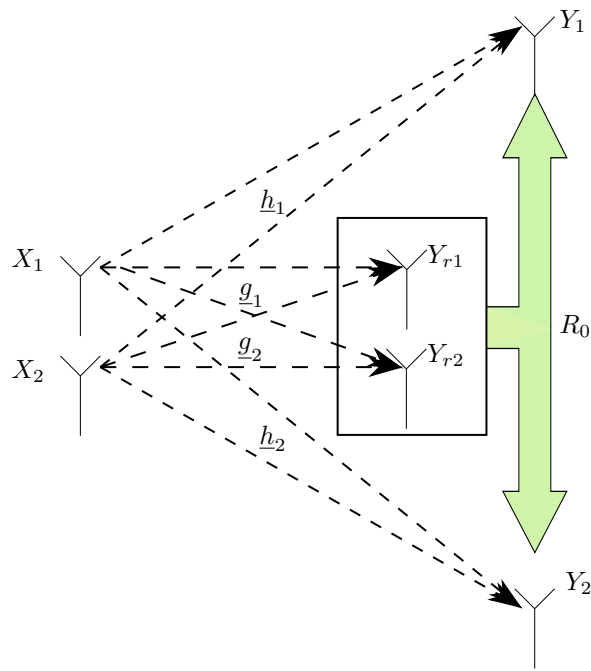

Fig. 1. A MIMO broadcast relay channel with a shared digital relay link of rate $R_0$ bits per channel use.

relay observes the destination noise:

$$Y = X + Z \tag{1a}$$
$$Y_r = Z. \tag{1b}$$

This channel was first introduced in [1], and it was shown that a hash-and-forward (HF) strategy, where the relay randomly bins $Y_r$ and forwards the bin index of rate $R_0$ to the destination, improves the achievable rate by $R_0$ bits[2]. In essence, the HF scheme is a quantize-and-forward strategy with joint decoding of the relay and source codewords.

When only an approximate, or "noisy" cross determinism exists, i.e., $Y_r \approx f(X, Y)$, the $R_0$ improvements of the QF scheme incurs a penalty, and the optimality of QF is not known in general. For some special cases, like a modulo-sum relay channel with a noisy cross-determinism, it has been shown that QF is indeed optimal, although the capacity is strictly below the cut-set bound [2]. For a Gaussian relay channel model described by (1), the optimality of QF is not known, although QF with Gaussian quantization yields the best-known achievable rate. Motivated by these results and by the more general approach of [3], we expect that QF provides a powerful and general technique that performs close to the optimal in many cases of interest.

---

[1]We notice here that channels that involve a noiseless common relay-to-destination(s) link of given rate $R_0$ may serve as useful models for out-of-band relay. For example, in a wireless scenario, the relay may receive on the same frequency of the source and destination, and forward its downstream message on a different band, which is modeled here as a perfect bit-pipe of fixed rate.

[2]This direct binning scheme [1] is applicable only with discrete finite-alphabets for $Y_r$. For continuous alphabets, HF can be applied to a high-resolution quantized version of $Y_r$.

While the noise-observing channel defined in (1) is not practically interesting by itself, there are a number of practically relevant multiuser channels where a noisy cross-determinism can be identified, where the relay essentially observes the interference incurred at the destination. For example, in an interference channel with a common relay, the relay observes the interfering signals which act as noise for the unintended receivers. Cross determinism can indeed be used in the Gaussian two-users interference channel with a common relay to design a universal strategy to assist both users simultaneously using a common out-of-band relay link [4], and leads to practical coding schemes for this channel [5].

An interesting scenario where cross determinism can be leveraged is the broadcast channel with a relay considered in the present work. Intuitively, in a broadcast channel the signal intended for one user appears as noise to the other user. Since the relay observes (a function of) the source signal, it essentially observes the interference noise incurred at each receiver. This paper explores the possibility of using this inherent noisy cross-determinism to design QF relay strategies for the Gaussian MIMO broadcast relay channel.

Our main tool is the achievable rate of the generalized hash-and-forward (GHF) strategy for a general relay channel, which is stated in the following theorems:

*Theorem 1 ( [4], [6]):* Consider a memoryless relay channel defined by $p(y, y_r|x)$, with a digital relay link of rate $R_0$ between the relay and the destination. A quantize-and-forward strategy where the relay quantizes its observation $Y_r$ into $\hat{Y}_r$ and forwards the index of the bin containing the quantization codeword at rate $R_0$ to the destination, achieves the following rate for $X, Y, Y_r, \hat{Y}_r \sim p(x)p(y, y_r|x)p(\hat{y}_r|y_r)$:

$$R \leq I(X;Y) + \min\{R_0, I(\hat{Y}_r; Y_r|Y)\} \\ - \min\{R_0, I(\hat{Y}_r; Y_r|X, Y)\}. \quad (2)$$

Notice that the penalty term $\min\{R_0, I(\hat{Y}_r; Y_r|X, Y)\}$ can be thought as the loss due to a "noisy" cross-determinism. In fact, this penalty vanishes when $Y_r = f(X, Y)$. The GHF strategy is a form of QF with a modified joint decoding strategy suitable in multiuser channels; see [4], where it is shown that GHF improves upon QF in an interference channel with a common relay.

To apply these results to broadcast channel, we also need the following extension of the above theorem to a relay channel with known side information at the source:

*Theorem 2:* Consider a memoryless relay channel with transmitter side information defined by $p(s)p(x|s)p(y, y_r|s, x)$ with a digital relay link of rate $R_0$ between the relay and the destination. A quantize-and-forward strategy where the relay quantizes its observation $Y_r$ into $\hat{Y}_r$ and forwards the index of the bin containing the quantization codeword at rate $R_0$ to the destination, achieves the following rate for $S, \tilde{U}, X, Y, Y_r, \hat{Y}_r \sim p(s)p(\tilde{u}|s)p(x|\tilde{u},s)p(y,y_r|x)p(\hat{y}_r|y_r)$.

$$R \leq I(\tilde{U};Y) - I(\tilde{U};S) + \min\{R_0, I(\hat{Y}_r;Y_r|Y)\} \\ - \min\{R_0, I(\hat{Y}_r;Y_r|\tilde{U},Y)\}. \quad (3)$$

The proof of the above theorem is based on combining the dirty paper coding strategy with the decoding strategy of the GHF relay scheme, and will be included in the complete version of this paper.

*B. Related Work*

The broadcast channel with a relay was first studied in [7], where a number of achievable rate regions are derived using combinations of block Markov coding, superposition coding, and dirty paper coding. Several subsequent papers have studied this channel and in some degraded cases, the capacity region was established using a stack of coding strategies, namely, message splitting, block Markov, superposition, Wyner-Ziv, and dirty paper coding; see for example [8] and references therein. Rather than employing all possible strategies to obtain the best possible performance, we pick the QF scheme, and restrict ourselves to optimize the quantization procedure in the high-SNR regime where the background noise is small. The main contribution of this paper is an intuitive technique to design the relay quantizer and the dirty-paper coding scheme at the source to asymptotically achieve the cut-set bound for individual rates, and the sum-capacity.

## II. MIMO BROADCAST CHANNEL WITH MULTIPLE-ANTENNA RELAY

Consider a two-user broadcast channel with two antennas at the source and one antenna at each destination, with a relay equipped with two receive antennas and a common noiseless broadcast link of fixed rate $R_0$ bits per channel use to the two destinations. The source-to-destinations and source-to-relay channels are defined by

$$\underline{Y} \triangleq \begin{bmatrix} Y_1 \\ Y_2 \end{bmatrix} = \begin{bmatrix} h_{11} & h_{12} \\ h_{21} & h_{22} \end{bmatrix}^* \begin{bmatrix} X_1 \\ X_2 \end{bmatrix} + \begin{bmatrix} Z_1 \\ Z_2 \end{bmatrix} \\ \triangleq \begin{bmatrix} \underline{h}_1^* \\ \underline{h}_2^* \end{bmatrix} \underline{X} + \underline{Z} \\ \triangleq H^* \underline{X} + \underline{Z}, \quad (4a)$$

and by

$$\underline{Y}_r \triangleq \begin{bmatrix} Y_{r1} \\ Y_{r2} \end{bmatrix} = \begin{bmatrix} g_{11} & g_{12} \\ g_{21} & g_{22} \end{bmatrix}^* \begin{bmatrix} X_1 \\ X_2 \end{bmatrix} + \begin{bmatrix} Z_{r1} \\ Z_{r2} \end{bmatrix} \\ \triangleq \begin{bmatrix} \underline{g}_1^* \\ \underline{g}_2^* \end{bmatrix} \underline{X} + \underline{Z}_r \\ \triangleq G^* \underline{X} + \underline{Z}_r. \quad (4b)$$

respectively, where $*$ denotes conjugate transpose. In the above model, $X_1, X_2$ denote the signals transmitted by the two source antennas, $Y_1, Y_2$ denote the received signals at user one and two, and $Y_{r1}, Y_{r2}$ represent the received signals at the two relay antennas. Upon observing $\underline{Y}_r$, the relay forms a message of rate $R_0$ and broadcasts it to the two destinations. Here, $Z_1, Z_2 \sim \mathcal{CN}(0, N)$ and $Z_{r1}, Z_{r2} \sim \mathcal{CN}(0, N)$ represent independent receiver noise terms, and the total source signal power is constrained to be less than $P$, although the presented strategy is still applicable with a per antenna power constraint.

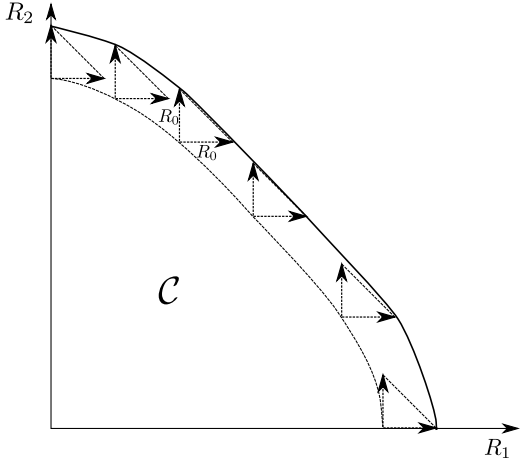

Fig. 2. The extended achievable rate region for the broadcast channel with a digital relay of rate $R_0$.

We refer to the asymptotically high-SNR regime as the regime where the noise variance $N$ vanishes, while all channel coefficients and the transmit power $P$ are fixed. Let $\mathcal{C}$ denotes the capacity region of the MIMO broadcast channel without the relay. Our goal in this paper is to show that, for the channel defined in (4), the rate region

$$(R_1, R_2) \in \Big\{(C_1 + \Delta R_1, C_2 + \Delta R_2) \big| (C_1, C_2) \in \mathcal{C},$$
$$\Delta R_1 + \Delta R_2 \leq R_0, \Delta R_1 \geq 0, \Delta R_2 \geq 0\Big\}, \quad (5)$$

can be achieved for asymptotically high SNR. This region is sketched in Fig. 2.

To prove the (asymptotic) achievability of the region in (5), we prove that for any rate pair $(C_1, C_2) \in \mathcal{C}$ achievable without using the relay, the rate pairs $(C_1 + R_0, C_2)$ and $(C_1, C_2 + R_0)$ are achievable when the relay is used, asymptotically for $N \to 0$.

## III. SUM RATE IMPROVEMENT

The Sato cut-set bound of [9] for the broadcast channel can be modified to include the relay. Using this bound, we can show that the sum of the two user rates $R_1$ and $R_2$ can be improved by at most $R_0$ using the relay. In this section, we wish to exploit cross-determinism to design a quantize-and-forward strategy to asymptotically achieve this bound, i.e., improving the sum rate by $R_0$ bits (per channel use).

The sum-capacity of the broadcast channel is achieved by using a two-stage dirty-paper encoding strategy along with optimal MMSE-DFE beam forming [10]–[12]. In this strategy, the source encodes the data for user one and user two using auxiliary random variables $U_1$ and $U_2$, and optimal beamforming vectors $\underline{v}_1$ and $\underline{v}_2$. The source transmit signal vector is given by:

$$\underline{X} = \underline{v}_1 U_1 + \underline{v}_2 U_2, \quad (6)$$

where $U_i \sim \mathcal{CN}(0, p_i)$, $i = 1, 2$, with $P = p_1 + p_2$, and $\underline{v}_1 \, \underline{v}_2$ are the optimal beamforming vectors [10]. For a given desired point on the capacity region boundary, the encoder picks an encoding order, choosing to encode the data for one user prior to the other. For the encoding order $U_1 \to U_2$, the source first encodes the data for user one via $U_1$, and then encodes the data for user two via $U_2$, by using dirty paper coding and treating the interference caused by $U_1$ at receiver two as known side information at the source encoder. At receiver one, the interference caused by $U_2$ is treated as noise.

The received signals at user one and two and the relay are given by:

$$Y_1 = (\underline{h}_1^* \underline{v}_1)U_1 + (\underline{h}_1^* \underline{v}_2)U_2 + Z_1 \quad (7)$$
$$Y_2 = (\underline{h}_2^* \underline{v}_1)U_1 + (\underline{h}_2^* \underline{v}_2)U_2 + Z_2 \quad (8)$$
$$\underline{Y}_r = (G^* \underline{v}_1)U_1 + (G^* \underline{v}_2)U_2 + \underline{Z}_r. \quad (9)$$

When $U_1$ is encoded first, the interference incurred to user one by $U_2$ is observed also by the relay. In other words, from user one's perspective, this scenario is an instance of the noise-observing relay channel discussed in Section I-A, asymptotically as $P$ tends to infinity. Thus, we expect that a simple quantize-and-forward strategy asymptotically improve the rate of user one by $R_0$ bits, for the encoding order $U_1 \to U_2$. Using the same strategy for the reverse encoding order $U_2 \to U_1$, the rate of user two can be improved by $R_0$ bits asymptotically, since receiver two treats the interference from $U_1$ as noise. Consequently, the sum capacity is asymptotically improved by $R_0$ bits by using the relay.

More specifically, the relay first forms the scalar observation $\tilde{Y}_r \triangleq \underline{w}_r^* \underline{Y}_r$ by projecting along the receive beamforming vector $\underline{w}_r$. Next, it quantizes $\tilde{Y}_r$ into $\hat{Y}_r$ using a Gaussian quantizer with MSE distortion $q$; in other words, we have $\tilde{Y}_r = \hat{Y}_r + \eta$, where $\eta \sim \mathcal{CN}(0, q)$.

A random bin index of rate $R_0$ for $\hat{Y}_r$ is sent over the digital relay link to the destinations. Assume without loss of generality that user one is encoded first. By Theorem 1, the rate improvement for user one using this relay strategy is given by:

$$\Delta R_1 = \min\{R_0, I(\hat{Y}_r; \tilde{Y}_r | Y_1)\} - \min\{R_0, I(\hat{Y}_r; \tilde{Y}_r | U_1, Y_1)\}.$$

Now, if we choose $\underline{w}_r$ and $q$, such that asymptotically

$$I(\hat{Y}_r; \tilde{Y}_r | Y_1) \geq R_0, \quad (10)$$

then the rate of user one is asymptotically improved by $R_0$. In fact, we have that given $U_1$ and $Y_1$, then $U_2$ is (asymptotically) deterministically fixed as $U_2 = (Y_1 - (\underline{h}_1^* \underline{v}_1)U_1)/(\underline{h}_1^* \underline{v}_2)$ as $N \to 0$, and thus $\tilde{Y}_r$ is also (asymptotically) deterministically fixed given $U_1$ and $Y_1$. Therefore, we have a vanishing penalty term $I(\hat{Y}_r; \tilde{Y}_r | U_1, Y_1) \to 0$ as $N \to 0$. In summary, by choosing the beamforming vectors $\underline{v}_1$ and $\underline{v}_2$ that achieve the capacity region without the relay, the sum capacity is asymptotically improved by $R_0$ by using a relay strategy that improves the rate of the *interfered user* (i.e., the user that is encoded first, in the successive encoding dirty-paper strategy).

### A. A Failing Strategy

Does the relay message simultaneously improve the rate of the user experiencing no interference (the user that is encoded second, in the successive dirty-paper strategy)? This would be against the cut-set bound for the sum rate if the rate of the non-interfered user also gets improved by the above relay scheme. It is however instructive to understand what happens to the non-interfered user rate.

Assume again that user one is encoded first. By Theorem 2, the rate improvement for user two using the above strategy is given by:

$$\Delta R_2 < \min\{R_0, I(\hat{Y}_r; \tilde{Y}_r | Y_2)\} - \min\{R_0, I(\hat{Y}_r; \tilde{Y}_r | \tilde{U}_2, Y_2)\}, \quad (11)$$

where

$$\tilde{U}_2 \triangleq k(\underline{h}_2^* \underline{v}_1) U_1 + (\underline{h}_2^* \underline{v}_2) U_2, \quad (12)$$

and

$$k = \frac{|\underline{h}_2^* \underline{v}_2|^2 p_2}{|\underline{h}_2^* \underline{v}_2|^2 p_2 + N}. \quad (13)$$

Here, $\tilde{U}_2$ denotes the dirty-paper coding auxiliary codebook variable and $\kappa$ is the dirty-paper scaling factor. Notice that from (8), for user two, the interference signal is given by $(\underline{h}_2^* \underline{v}_1) U_1$ and the effective receiver SNR is given by $|\underline{h}_2^* \underline{v}_2|^2 p_2 / N$. In Costa's notation [13], $U_1$ corresponds to $S$, $U_2$ corresponds to $X$, and $\tilde{U}_2$ corresponds to $U$.

Asymptotically as $N \to 0$, we have $k \to 1$ and, consequently, $\tilde{U}_2 \to Y_2$ by (8). Thus, $\Delta R_2 \to 0$ asymptotically by (11), since

$$I(\hat{Y}_r; \tilde{Y}_r | \tilde{U}_2, Y_2) \to I(\hat{Y}_r; \tilde{Y}_r | Y_2).$$

This result is consistent with the cut-set bound. However, it raises the question of what relay strategy would improve the rate of the non-interfered user. It is crucial to answer this question in order to devise a relay strategy that asymptotically achieves the rate region in (5), since the optimal dirty-paper encoding order is dictated by the rate point on the boundary of $\mathcal{C}$ at which we wish to operate [14]. As mentioned earlier, we need relay strategies to asymptotically achieve both rate pairs $(C_1, C_2 + R_0)$ and $(C_1 + R_0, C_2)$, for any point $(C_1, C_2)$ on the boundary, and not just one of the two points depending on the encoding order.

To design a relay strategy capable of improving the rate of the non-interfered user, we consider a particular class of deterministic broadcast channel with a common relay in the next section. The capacity region of the deterministic broadcast channel is known to be achievable using a simple dirty-paper coding strategy. However, the addition of a relay requires a more complicated encoding strategy involving additional auxiliary random variables. The encoding strategy for the deterministic channel serves us as a guide to devise a relay strategy for the Gaussian MIMO broadcast channel with a common relay.

## IV. DETERMINISTIC BROADCAST CHANNELS WITH A DIGITAL RELAY LINK

Consider a memoryless deterministic broadcast relay channel where

$$Y_1 = f_1(X) \quad (14)$$
$$Y_2 = f_2(X), \quad (15)$$

with a relay observing

$$Y_r = f_r(X), \quad (16)$$

where $f_1$, $f_2$ and $f_r$ are deterministic functions, and $X \in \mathcal{X}$, where $\mathcal{X}$ is a finite set of channel input alphabets. The relay is equipped with a noiseless broadcast link of rate $R_0$, bits per channel use, to the two destinations.

The capacity region of the discrete deterministic broadcast channel without the relay is known [15], and is given by:

$$R_1 \leq H(Y_1) \quad (17a)$$
$$R_2 \leq H(Y_2) \quad (17b)$$
$$R_1 + R_2 \leq H(Y_1, Y_2), \quad (17c)$$

and it is achievable using Marton coding strategy. Recall that the restricted Marton region for the broadcast channel without common message is given by:

$$R_1 \leq I(U; Y_1) \quad (18a)$$
$$R_2 \leq I(V; Y_2) \quad (18b)$$
$$R_1 + R_2 \leq I(U; Y_1) + I(V; Y_2) - I(U; V), \quad (18c)$$

which is equivalent to (17) if we choose $U = Y_1$, and $V = Y_2$.

Now, we consider the relay observing $Y_r = f_r(X)$ with a common digital link of rate $R_0$ to the two users. The following theorem gives a cut-set outer bound for the achievable rate region:

*Theorem 3:* The capacity region of the discrete broadcast relay channel defined in (17) is included in the following region:

$$R_1 \leq H(Y_1) + R_0 \quad (19a)$$
$$R_2 \leq H(Y_2) + R_0 \quad (19b)$$
$$R_1 + R_2 \leq H(Y_1 Y_2) + R_0. \quad (19c)$$

The proof is based on using the cut-set bound and will be included in the complete version of this paper.

In the following, we show that the above outer bound is indeed achievable when

$$H(Y_r | Y_1 Y_2) \geq R_0, \quad (20)$$

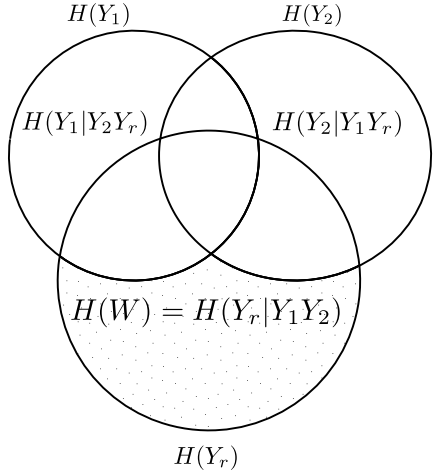

Fig. 3. A symbolic Ven diagram describing the overlaps between random variables $X, Y_1, Y_2, Y_r$.

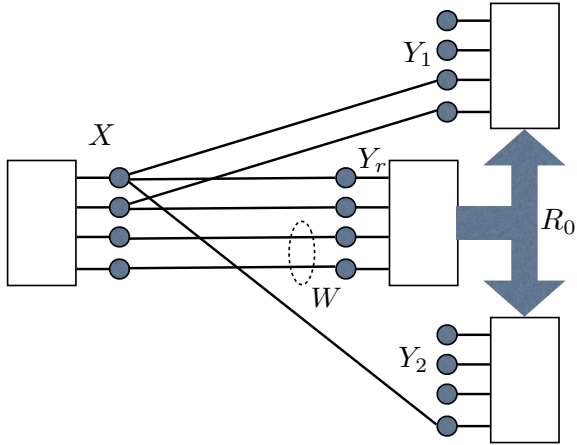

Fig. 4. A linear deterministic broadcast-relay channel satisfying the assumptions of Section IV.

and there is a function $g$ such that $W = g(Y_r)$ is independent of $Y_1, Y_2$ with $H(W) = H(Y_r|Y_1Y_2)$. An intuition for such situation is provided by the Venn diagram representing entropies in Fig. 3. In short, we assume that there exist a deterministic mapping $g(\cdot)$ that extracts the part of $Y_r$ that is independent of both $Y_1$ and $Y_2$. A simple example of such a situation, included here with the unique purpose to show that indeed the set of channels considered in this section is non-trivial (non-empty) is based on the linear deterministic model of [3] and shown in Fig. 4. In this example, we have $Y_r = X$ and $W$ corresponds to the two least-significant bits of $Y_r$. We have $H(W) = H(Y_r|Y_1, Y_2) = 2$ and clearly $W$ is independent of $Y_1$ and $Y_2$ by choosing $X \sim$ Uniform.

We begin by deriving an achievable rate region for a general discrete memoryless broadcast relay channel:

*Theorem 4:* Consider a two-user memoryless broadcast relay channel defined by $p(y_1, y_2, y_r|x)$, where $X$ denotes the source signal, and $Y_1, Y_2, Y_r$ denote the received signals at user one, user two, and the relay, respectively. The relay sends a common message to both users over a shared digital (noiseless) link of rate $R_0$ bits per channel use (see 1). For this channel, the following rate region is achievable:

$$R_1 \leq I(U;Y_1) + \min\{R_0, I(\hat{Y}_r; Y_r|Y_1)\} \quad (21a)$$
$$\quad - \min\{R_0, I(\hat{Y}_r; Y_r|U, Y_1)\} \triangleq I(U;Y_1) + \Delta R_1$$
$$R_2 \leq I(V;Y_2) + \min\{R_0, I(\hat{Y}_r; Y_r|Y_2)\} \quad (21b)$$
$$\quad - \min\{R_0, I(\hat{Y}_r; Y_r|V, Y_2)\} \triangleq I(V;Y_2) + \Delta R_2$$
$$R_1 + R_2 \leq I(U;Y_1) + I(V;Y_2) - I(U;V)$$
$$\quad + \Delta R_1 + \Delta R_2, \quad (21c)$$

for $U, V, X, Y_1, Y_2, Y_r, \hat{Y}_r \sim p(u,v)p(x|u,v)p(y_1,y_2,y_r|x)p(\hat{y}_r|y_r)$.

*Proof:*

*Encoding:* The source employs random coding and joint binning similar to the Marton's coding strategy. Fix a probability distribution $p(u,v)$. The source generates two sets of i.i.d.-generated codewords $U^n(i)$ and $V^n(j)$ of length $n$, for $i = 1, \ldots, 2^{n\tilde{R}_1}$, and $j = 1, \ldots, 2^{n\tilde{R}_2}$, according to probability distributions $p(u)$ and $p(v)$, respectively. Bin the set of $(i,j)$ pairs randomly and uniformly into $2^{n(R_1+R_2)}$ bins, labeled by $(k,l)$ for $k = 1, \ldots, 2^{nR_1}$ and $l = 1, \ldots, 2^{nR_2}$. To send the pair of messages $(k,l)$, find a pair of jointly typical codewords $(\mathbf{u}(i), \mathbf{v}(j))$ in the bin labeled by $(k,l)$. Then, find an $\mathbf{x}(k,l)$ jointly typical with $(\mathbf{u}(i), \mathbf{v}(j))$ to encode the pair of messages $(k,l)$.

This encoding strategy is successful if

$$R_1 + R_2 \leq \tilde{R}_1 + \tilde{R}_2 - I(U;V), \quad (22)$$

which ensures that in every bin, a pair of jointly typical $(U^n(i), V^n(j))$ codewords is always found [16].

A GHF strategy is used at the relay. The relay quantizes its observation $Y_r$ into $\hat{Y}_r$ and sends a bin index of rate $R_0$ for $\hat{Y}_r$ to the two destinations.

*Decoding:* User one decodes $U^n(i)$, $i = 1, \ldots, 2^{n\tilde{R}_1}$, using the decoding strategy of GHF. Likewise, user two decodes $V^n(j)$, $j = 1, \ldots, 2^{n\tilde{R}_2}$ using the decoding strategy of GHF. By Theorem 1, decoding is successful with arbitrarily high probability for sufficiently large $n$ if:

$$\tilde{R}_1 \leq I(U;Y_1) + \min\{R_0, I(\hat{Y}_r; Y_r|Y_1)\}$$
$$\quad - \min\{R_0, I(\hat{Y}_r; Y_r|U, Y_1)\}, \quad (23a)$$
$$\tilde{R}_2 \leq I(V;Y_2) + \min\{R_0, I(\hat{Y}_r; Y_r|Y_2)\}$$
$$\quad - \min\{R_0, I(\hat{Y}_r; Y_r|V, Y_2)\}. \quad (23b)$$

Combining (22) and (23) results in the achievable rate region of (21), since $R_1 \leq \tilde{R}_1$ and $R_2 \leq \tilde{R}_2$. ∎

### A. Achieving the Capacity Region

Our next goal is to appropriately choose $U, V$ such that (21) coincides with (19). Surprisingly, the optimal choice for $(U, V)$ is not trivial in presence of the relay. Although choosing $U = Y_1$ and $V = Y_2$ is optimal for the deterministic broadcast channel without the relay, $(U, V) = (Y_1, Y_2)$ results in $\Delta R_1 = \Delta R_2 = 0$ in (21).

In our case, by assumption, we have that $W = g(Y_r)$ is a deterministic function of $Y_r$, independent of $Y_1$ and $Y_2$, with $H(W) = H(Y_r|Y_1Y_2)$. Consider the choice of auxiliary random variables $U = Y_1W$, and $V = Y_2$, and choose the relay quantization variable as $\hat{Y}_r = W$. Then, we have:

$$\begin{aligned}\Delta R_1 &= \min\{R_0, I(W; Y_r|Y_1)\} \\ &\quad - \min\{R_0, I(W; Y_r|Y_1W, Y_1)\} \\ &= \min\{R_0, H(W)\} - 0 \\ &= \min\{R_0, H(Y_r|Y_1Y_2)\} \\ &= R_0,\end{aligned}$$

and,

$$I(Y_1W; Y_1) = H(Y_1).$$

where we used assumption (20). Thus, for $U = Y_1W$, $V = Y_2$, and $\hat{Y}_r = W$, the region of (21) reduces to:

$$\begin{aligned}R_1 &\leq H(Y_1) + R_0 \\ R_2 &\leq H(Y_2) \\ R_1 + R_2 &\leq I(Y_1W; Y_1) + I(Y_2; Y_2) - I(WY_1; Y_2) + R_0 \\ &\leq H(Y_1) + H(Y_2) - I(Y_1; Y_2) - I(W; Y_2|Y_1) + R_0 \\ &\leq H(Y_1Y_2) - H(W|Y_1) + H(W|Y_1Y_2) + R_0 \\ &\leq H(Y_1Y_2) - H(W) + H(W) + R_0 \\ &\leq H(Y_1Y_2) + R_0. \quad (24)\end{aligned}$$

Likewise, by choosing $U = Y_1$, $V = Y_2W$, and $\hat{Y}_r = W$, (21) reduces to:

$$\begin{aligned}R_1 &\leq H(Y_1) \\ R_2 &\leq H(Y_2) + R_0 \\ R_1 + R_2 &\leq H(Y_1Y_2) + R_0. \quad (25)\end{aligned}$$

By time sharing between the two strategies, we get the region given by the outer bound in (19).

### B. Discussion

Let's focus on the relay strategy that improves the rate of user one. The optimality of the above relay strategy is a consequence of:

$$I(\hat{Y}_r; Y_r|U, Y_1) = 0 \quad (26)$$

for $U = WY_1$ and $\hat{Y}_r = W$. This is again a manifestation of the cross determinism: the relay forms a function of $Y_r$, namely $\hat{Y}_r$, which is deterministically fixed given the source signal $U$ and the observed signal $Y_1$ at receiver one. We can further check that if the relay simply chooses $\hat{Y}_r = Y_r$ (like the HF quantizer for (1)), the quantization penalty term $I(\hat{Y}_r; Y_r|U, Y_1)$ is nonzero; unless $U = Y_1Y_r$, which results in reduced rate for user two, since for $V = Y_2$ we have (Theorem 4):

$$\begin{aligned}R_1 + R_2 &\leq H(Y_1) + H(Y_2) - I(Y_1Y_r; Y_2) + R_0 \\ &= H(Y_1Y_2) - I(Y_r; Y_2|Y_1) + R_0 \quad (27)\end{aligned}$$

The penalty term $I(Y_r; Y_2|Y_1)$ in (27) is caused by the "overlap" between $Y_r$ and $Y_2$ (see Figs. 3 and 4). This may illustrate why the rate improvement for the non-interfered user tends to zero asymptotically in the previous relay strategy for the Gaussian MIMO broadcast channel: The potential interference between $\hat{Y}_r = \underline{w}_r^* \underline{Y}_r$ and $Y_2$ results in reduced rate for user two. As we see from the deterministic channel example, the relay quantizer should be carefully chosen in a way to minimize overlaps. For the Gaussian MIMO channel, this suggests a careful choice of $\underline{w}_r$ and the use of an additional auxiliary encoding random variable for dirty-paper coding in order to have the relay improve the rate of the non-interfered user.

## V. IMPROVING THE RATE OF THE NON-INTERFERED USER

We observed in the Section III that a two-stage dirty paper encoding fails to achieve the asymptotic capacity. As we shall see, a successful high-SNR achievability strategy in order to improve the rate of the non-interfered user consists of a three-stage dirty paper encoding strategy with three auxiliary random variables involved, inspired by the deterministic model treated in the previous section. Further, the relay no longer directly quantizes its observation, but a special projection at the relay is required, reminiscent of the relay quantization strategy for the deterministic broadcast relay channel, where the relay quantization ($W$ in Fig. 3) avoids overlaps with the user signals.

### A. Encoding

Without loss of generality, let's focus on a point $(C_1, C_2) \in \mathcal{C}$ achieved by successive encoding order $U_1 \to U_2$ (i.e., user one first). Our goal here is to find a strategy that improves asymptotically the rate of user two by $R_0$. The source employs a three-stage dirty-paper coding strategy along with beamforming. The message of user two is split in two parts, that will be denoted as two virtual users two and three. Successive encoding is used in the order $U_1 \to U_2 \to U_3$. First, the message of user one is encoded via $U_1$, while interference caused by $U_2$ and $U_3$ is treated as noise at receiver one. Next, the first part of user two's message is encoded via $U_2$, while the interference caused by $U_1$ at receiver two is treated as known side information at the encoder. Finally, the second part of user two's message is encoded via $U_3$, while the interference caused by $U_1$ and $U_2$ are treated as known side information.

The transmit signal vector is given by

$$\underline{X} = U_1 \cdot \underline{v}_1 + U_2 \cdot \underline{v}_2 + U_3 \cdot \underline{v}_3, \quad (28)$$

where $U_i \sim \mathcal{CN}(0, p_i)$, $i = 1, 2, 3$, with $P = p_1 + p_2 + p_3$, and $\underline{v}_1$ and $\underline{v}_2 = \underline{v}_3$ are the optimal (MMSE-DFE) beamforming vectors [10]. It is straightforward to check that when the optimal beamforming vectors are used, this three-stage dirty paper coding strategy still achieves the same point on $\mathcal{C}$ that is achieved by the classical two-stage encoding. The resulting

received signals are given by:

$$\underline{Y} = \begin{bmatrix} \underline{h}_1^* \underline{v}_1 & \underline{h}_1^* \underline{v}_2 & \underline{h}_1^* \underline{v}_3 \\ \underline{h}_2^* \underline{v}_1 & \underline{h}_2^* \underline{v}_2 & \underline{h}_2^* \underline{v}_3 \end{bmatrix} \begin{bmatrix} U_1 \\ U_2 \\ U_3 \end{bmatrix} + \begin{bmatrix} Z_1 \\ Z_2 \end{bmatrix} \quad (29)$$

$$\triangleq \begin{bmatrix} \tilde{\underline{h}}_1^* \\ \tilde{\underline{h}}_2^* \end{bmatrix} \underline{U} + \underline{Z}, \quad (30)$$

and at the relay,

$$\underline{Y}_r = \underline{G}^* \underline{v}_1 \cdot U_1 + \underline{G}^* \underline{v}_2 \cdot U_2 + \underline{G}^* \underline{v}_3 \cdot U_3 + \underline{Z}_r \quad (31)$$

$$\triangleq \begin{bmatrix} \tilde{\underline{g}}_1^* \\ \tilde{\underline{g}}_2^* \end{bmatrix} \underline{U} + \underline{Z}_r \quad (32)$$

$$\triangleq \tilde{G}^* \underline{U} + Z_r. \quad (33)$$

We shall provide an asymptotic (for high SNR) rate increase of $R_0$ to user two, by improving the rate of the virtual user $U_2$. To achieve this goal, we seek an asymptotic cross-determinism for this virtual user. The dirty paper encoding for $U_2$ with known side information $(\underline{h}_2^* \underline{v}_1) U_1$ involves the auxiliary random variable (see [13]):

$$\tilde{U}_2 \triangleq (\underline{h}_2^* \underline{v}_2) U_2 + k(\underline{h}_2^* \underline{v}_1) U_1, \quad (34)$$

where,

$$k = \frac{|\underline{h}_2^* \underline{v}_2|^2 p_2}{|\underline{h}_2^* \underline{v}_2|^2 p_2 + |\underline{h}_2^* \underline{v}_3|^2 p_3 + N}, \quad (35)$$

since $(\underline{h}_2^* \underline{v}_3) U_3$ is treated as noise along $Z_2$. Similarly, dirty paper encoding of $U_3$ is performed using the auxiliary random variable:

$$\tilde{U}_3 \triangleq (\underline{h}_2^* \underline{v}_3) U_3 + l\Big( (\underline{h}_2^* \underline{v}_2) U_2 + (\underline{h}_2^* \underline{v}_1) U_1 \Big), \quad (36)$$

where,

$$l = \frac{|\underline{h}_2^* \underline{v}_3|^2 p_3}{|\underline{h}_2^* \underline{v}_3|^2 p_3 + N}. \quad (37)$$

Notice that asymptotically as $N \to 0$, we have $l \to 1$ and $\tilde{U}_3 \to Y_2$. Thus, we do not expect $(\tilde{U}_3, Y_2)$ to form an asymptotic cross-determinism for user two.

However, we will see that as $N \to 0$, $(\tilde{U}_2, Y_2)$ can deterministically reveal a linear function of $\underline{Y}_r$. We exploit this cross-determinism to improve the rate of user two by $R_0$ bits, in the high-SNR regime.

To design the relay quantizer, we utilize an asymptotic linear cross-deterministic relation between $Y_r$ and $(\tilde{U}_2, Y_2)$. First, identify a receive beamforming vector $\underline{w}_r$ at the relay, and a virtual beamforming vector $\underline{w}_2$ at user two, such that

$$\underline{w}_r^* \underline{Y}_r = \underline{w}_2^* \begin{bmatrix} Y_2 \\ \tilde{U}_2 \end{bmatrix}, \quad (38)$$

asymptotically as $N \to 0$, for all $U_1, U_2, U_3$. Next, the relay quantizes $\tilde{Y}_r \triangleq \underline{w}_r^* \underline{Y}_r$ into $\hat{Y}_r$ using a Gaussian quantizer with distortion $q$.

Now, from Theorem 2, if we choose $q > 0$ such that asymptotically

$$I(\hat{Y}_r; \tilde{Y}_r | Y_2) = R_0, \quad (39)$$

then, the rate of the message encoded by $U_2$ (and, hence, the rate of user two), is asymptotically improved by $R_0$. The reason lies in the asymptotic cross-determinism of $\tilde{Y}_r$ in terms of $\tilde{U}_2$ and $Y_2$, which results in an asymptotically vanishing quantization penalty:

$$I(\hat{Y}_r; \tilde{Y}_r | \tilde{U}_2, Y_2) \to 0, \quad (40)$$

since by (38), $\tilde{Y}_r$ is deterministically fixed given $\tilde{U}_2$ and $Y_2$, asymptotically.

It remains to find $\underline{w}_r$ and $\underline{w}_2$ for (38) to hold asymptotically. Neglecting the noise, (38) can be written as

$$\underline{w}_r^* \tilde{G}^* \underline{U} = \underline{w}_2^* \begin{bmatrix} \tilde{\underline{h}}_2^* \\ \hat{\underline{h}}_2^* \end{bmatrix} \underline{U}, \quad (41)$$

where,

$$\hat{\underline{h}}_2 \triangleq \begin{bmatrix} k \cdot \underline{v}_1^* \underline{h}_2 \\ \underline{v}_2^* \underline{h}_2 \\ 0 \end{bmatrix},$$

with $\tilde{U}_2 = \hat{\underline{h}}_2^* \underline{U}$ (see (34)).

Now, (41) holds for all $U_1, U_2, U_3$, if

$$\tilde{G} \underline{w}_r = \begin{bmatrix} \tilde{\underline{h}}_2 & \hat{\underline{h}}_2 \end{bmatrix} \underline{w}_2. \quad (42)$$

The vectors $\underline{w}_r$ and $\underline{w}_2$ can be found by solving:

$$\begin{bmatrix} \tilde{\underline{h}}_2 & \hat{\underline{h}}_2 & -\tilde{G} \end{bmatrix} \begin{bmatrix} \underline{w}_2 \\ \underline{w}_r \end{bmatrix} = 0, \quad (43)$$

which is an under-determined linear system and always has a nonzero solution. Since we are interested in a nonzero solution for $\underline{w}_r$, let $\underline{w}_r = [w_{r1}; 1]$ or $\underline{w}_r = [1; w_{r2}]$. Then, an explicit solution for $\underline{w}_r$ and $\underline{w}_2$ can be found by solving the linear set of equations:

$$\begin{bmatrix} \tilde{\underline{h}}_2 & \hat{\underline{h}}_2 & -\tilde{\underline{g}}_1 \end{bmatrix} \begin{bmatrix} \underline{w}_2 \\ w_{r1} \end{bmatrix} = \tilde{\underline{g}}_2, \quad (44)$$

or,

$$\begin{bmatrix} \tilde{\underline{h}}_2 & \hat{\underline{h}}_2 & -\tilde{\underline{g}}_2 \end{bmatrix} \begin{bmatrix} \underline{w}_2 \\ w_{r2} \end{bmatrix} = \tilde{\underline{g}}_1. \quad (45)$$

Thus, a sufficient condition to have a nonzero solution for $\underline{w}_r$ is that either

$$\begin{bmatrix} \tilde{\underline{h}}_2 & \hat{\underline{h}}_2 & -\tilde{\underline{g}}_1 \end{bmatrix} \quad (46)$$

or

$$\begin{bmatrix} \tilde{\underline{h}}_2 & \hat{\underline{h}}_2 & -\tilde{\underline{g}}_2 \end{bmatrix} \quad (47)$$

is full rank. Now, since $\hat{\underline{h}}_2$ can be tweaked via controlling $k$ in (35) by choosing an appropriate power splitting ratio between $p_2$ and $p_3$, we only need either

$$\begin{bmatrix} \tilde{\underline{h}}_2 & \tilde{\underline{g}}_1 \end{bmatrix}$$

or

$$\begin{bmatrix} \tilde{\underline{h}}_2 & \tilde{\underline{g}}_2 \end{bmatrix}$$

to be full rank for a nonzero solution to exist. In other words, in the equivalent channel (29), at least one relay antenna should receive an independent observation of the user two's signal, which is guaranteed to happen with probability one in a fading scenario when channel coefficients are independently distributed.

## VI. CONCLUSION

We introduced a technique for designing the relay quantizer in the quantize-and-forward relay strategy for Gaussian MIMO broadcast channel with a multiple antenna relay. The main idea is to leverage a noisy cross-deterministic relation in order to make the penalty for quantization vanish in the high-SNR regime. For the MIMO broadcast channel with a multiple-antenna relay, forwarding the same relay message to the users via a common out-of-band (digital) link of fixed rate $R_0$, the source message intended for one user interferes with the other user. By observing a noisy linear projection of the source signal, the relay essentially observes the interference appeared as noise to the unintended user. A cross-deterministic relation can be identified based on this observation and used to design appropriate encoding and quantization schemes to asymptotically achieve the cut-set bounds for individual rates, as well as the sum rate for the Gaussian MIMO broadcast channel with multiple-antenna relay. We used the intuition driven by a deterministic broadcast channel model with a relay as a simplified problem and introduced a capacity-achieving strategy for the deterministic case under some conditions. The encoding strategy for the deterministic case suggests a non-trivial three-stage dirty paper coding encoding along with a receiver beamforming strategy at the relay for the Gaussian channel, that would have been difficult to guess from first principles.